\documentclass{elsart}

\usepackage{epsfig}

\usepackage{amssymb}
\begin{document}

\begin{frontmatter}



\title{Cosmogenesis Backgrounds, Experiment Depth and
the Solar Neutrino TPC}

\author{G. Bonvicini},
\author{A. Schreiner\corauthref{cor1}}
\address{Wayne State University, Detroit MI 48201}
\corauth[cor1]{Corresponding author. Permanent address: Physics Dept., Wayne State University, Detroit MI 48201, USA. Tel.: +1-313-577-5067; Fax: +1-313-577-3932. \ead{schrein@physics.wayne.edu}}

\begin{abstract}
A Time Projection Chamber (TPC) is one of the promising candidates
to perform unique measurements in solar neutrino physics. Its
features will enable it to work at depths of the order of 2000
mwe. This paper describes an estimation of the expected cosmogenic
background at different depths including also the background due
to fission activation of the TPC material above ground.

\end{abstract}

\end{frontmatter}

\section{Introduction}
A solar neutrino Time Projection Chamber (TPC) provides unique and
very powerful information about solar neutrinos, their nuclear
parents and their fluxes (electron and non-electron), which
ultimately translates into better measurements of the neutrino
mass and mixing parameters \cite{snowmass}.

The TPC measures only $\nu -e$ elastic scattering, and determines the
electron direction and kinetic energy. This information allows reconstruction
of the incident neutrino energy.

Directionality also allows to use only electrons recoiling away
from the sun, effectively eliminating most background events. The
rest of the background events is eliminated by subtracting a
12-hours delayed coincidence, very much the same way as the
SuperKamiokande experiment can extract a clean solar neutrino
sample mixed with a much larger non-directional background
\cite{superk}. In that experiment, however, low recoil angle and
large multiple scattering conspire to destroy any directional
information that could help to determine the neutrino energy. The
TPC considered here looks at events with large scattering angle
and low multiple scattering \cite{snowmass}.

This paper is written for the sole purpose of confirming that this
detector can work at moderate depths. This has been the subject of
much skepticism in public meetings, and besides the main results,
consistency checks are offered. There are several advantages in
working at moderate depths:
\begin{itemize}
\item Virtually all available underground sites can accommodate this detector,
allowing us to concentrate on the aspects that really matter: low excavation
costs, low radon, quality and quantity of dust contamination,
and strong laboratory support.
\item This detector will be housed in a gigantic hall. Low overburden reduces
the civil engineering complexity.
\item Very deep sites have unacceptable rock temperatures, making
the assembly and operation more difficult and expensive.
\item Moderate depths allow the plentiful observation of certain background events,
which provide excellent calibration of the TPC performance. These
include \cite{position}:
\begin{itemize}
\item $O(10^6)$ straight, minimum-ionizing cosmic rays per year,
which allow a complete calibration of the electric field, gain,
electronics amplification, electron lifetime, and diffusion,
without a need for separate hardware and software. A TPC ``without
moving parts'' minimizes failures and avoids costly and
potentially contaminating accesses.
\item $O(10^5)$ delta rays above 100 keV per year. These have a well-known
kinematic relation between the angle and energy of the delta ray, providing
free and very accurate calibration of the detector resolution.
\end{itemize}
\end{itemize}

In section \ref{intr}, we introduce concepts useful to the
analysis. The main results are presented in section \ref{resul},
divided into analyzes of cosmogenesis backgrounds generated above
and below ground. Depths between 1500 and 3000 mwe are considered.
In section \ref{x-check}, we cross check our results against
other, published results and draw the conclusions.

\section{Overview of the TPC and cosmogenesis}\label{intr}
\begin{figure}[pthb]
\centerline{\psfig{file=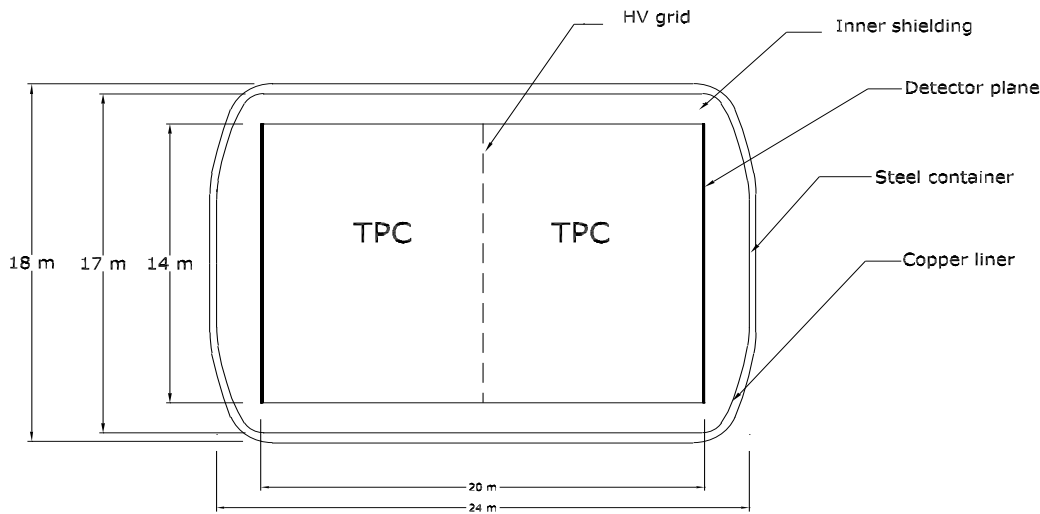,height=3.5in,width=5.5in,angle=0}}
     \caption{Sketch of TPC cage and surroundings.}
     \label{drawi}
\end{figure}

In the following, we consider a TPC whose parameters have been
outlined in \cite{position}. Briefly, as shown in Fig.\ref{drawi},
it is a TPC of 4000 cubic meters volume, surrounded by a 1.5
meters of radiopure, hydrogen rich material (either plastic or
water ice), which forms the inner shielding. Surrounding the inner
shielding are 30 centimeters of steel, forming the pressure vessel
and outer shielding. The weight of the inner shield is 1.9
kilotons, and the weight of the iron is 3.6 kilotons. The last
five centimeter thick layer of the inner shield, surrounding the
TPC, weighs 59 tons. We mention it here because it is particularly
helpful for instantly estimating self-shielded backgrounds (see
section 2).

The TPC contains 7.6 tons of gas at ten atmospheres, and the gas
is a mixture of helium (97\%) and methane (3\%), for a total
methane weight of 840 kg. The TPC will also contain approximately
100 kg of electrodeposited copper and 2 kg of sense wires. With
these conditions, 14 meters of gas (the diameter of the chamber)
represent approximately 0.14 Compton interaction lengths,
indicating that roughly one in seven 1 MeV gamma rays entering the
chamber will interact with the gas. 

We have discussed elsewhere
\cite{position,elliott} that recoiling electrons will generate a
track  of 9 cm length for 100 keV electrons and of 1.8 m for 600
keV electrons. The maximum drift time is 8.3 milliseconds, so that
the TPC snaps an electronic picture of the gas volume 120 times
per second. Whenever an event is encountered, a time interval
around the event of $\pm 9$msec is considered for vetoing.

The signal (a single electron) is readily simulated by $\beta$
radiation inside the active volume, and $\gamma$ radiation from
any source, generating a Compton scattering inside the chamber.

When looking at the volume distribution, the TPC occupies about
62\%  of the volume in Fig.\ref{drawi}. Most cosmic rays will
track through the sensitive volume and provide a direct veto that
eliminates all short-lived radionuclides as well as neutrons,
which thermalize and are absorbed in about 0.1 msec in the inner
shield. About 25\%  of cosmic rays crossing the steel shell will
not cross the active volume and all types of radioactive
backgrounds produced by them will have to be considered.

When looking at the weight distribution, however, the active volume of the TPC
only has a 0.13\% share of the total weight. Cosmogenesis backgrounds produced
outside the active volume have a very small chance of interacting with the
gas.

In \cite{snowmass}, it was found that background rates of around
600 events per day, above 100 keV and before directional cuts,
result in a non-dominant background subtraction error over two
years of data taking. Over a longer period of time, even
background rates at the level of one thousand events per day will
not drastically affect the final precision of the experiment.

\subsection{Cosmic generation of radionuclides}
At any given depth, the quantity of a particular long-lived daughter $i$,
generated in a given sample of mother nuclei $j$, can be described by a
charging capacitor law
\begin{equation}
N_c(t)=\epsilon_{ij}(h)\tau(1-e^{-t/\tau})
\end{equation}
where $\tau$ is the daughter nuclide lifetime. The variable
$\epsilon_{ij}(h)$ carries all the information about the reaction
probability,
\begin{equation}
\epsilon_{ij}(h)=<\sigma_{ij}(h)>N_t\Phi(h).
\end{equation}
Here $N_t$ is the number of target nuclei, $\Phi(h)$ the
depth-dependent flux, and $<\sigma_{ij}>$ the cross-section for
the process, averaged over the depth-dependent spectrum,
\begin{equation}
<\sigma_{ij}(h)>={1\over\Phi(h)}\int dE\sigma_{ij}(E){d\Phi(E,h)\over dE}.
\end{equation}
 The values of $\epsilon_{ij}(h)$ for dominant reactions are
 given in table \ref{reac-prob}.

The cross section depends strongly on the depth. Besides the
overburden  attenuation, nuclide activation is dominated by
hadronic showers at sea level ($h=10.33$ meters of water
equivalent, mwe), by muon capture at intermediate depths and by
fast muons at all depths below 100 mwe \cite{nolte1}.

The activity during charge-up is
\begin{equation}
A_c(t)=\epsilon_{ij}(h)(1-e^{-t/\tau}).
\end{equation}
and after a long time one obtains:
\begin{equation}
N_c(\infty)=\epsilon_{ij}(h)\tau,\;\;\;A_c(\infty)=\epsilon_{ij}.
\end{equation}
For a sample that is charged up at the surface and taken deep
underground, the discharging capacitor equations hold:
\begin{equation}
N_d(t)=N_0e^{-t/\tau},\;\;\;A_d(t)={N_0\over\tau}e^{-t/\tau}.
\end{equation}
$N_0$ is, in all practical cases, close to
$N_c(\infty)$. It can be reduced by up to two orders of magnitude
for materials which are mined or taken deep underground, and are exposed
to surface radiation during transportation or processing before
storing underground again.

Cosmic fluxes, their energy and angular dependence were kindly
provided by E. Nolte \cite{nolte1}. Dr. Nolte provided us also
with the measured cross sections for nuclide production by fast muons, and
the measured muon capture cross sections to estimate this source
of backgrounds as well \cite{nolte1}.
\begin{table}
\begin{center}
\begin{tabular}{|c|c|c|c|}
 \hline
reaction    &   $\epsilon_{ij}$(0mwe),   &
$\epsilon_{ij}$(1500mwe), & $\epsilon_{ij}$(3000mwe),\\
&ton$^{-1}$day$^{-1}$&ton$^{-1}$day$^{-1}$&ton$^{-1}$day$^{-1}$\\\hline\hline
Fe$\rightarrow^{54}$Mn &   410000  &   15  &   0.73
\\\hline Cu$\rightarrow^{60}$Co &   91000  &   3.2    &   0.16
\\\hline C$\rightarrow^{7}$Be  &   19000   &   0.61    &   0.03
\\\hline C$\rightarrow^{11}$C  &   -   &   2.4    &   0.12\\\hline
\end{tabular}
\end{center}
\caption{The activation probabilities $\epsilon_{ij}(h)$, for significant
reactions and at various depths.\label{reac-prob}}
\end{table}

\subsection{Cross section approximations}
The following approximations were used when parameterizing nuclear
cross sections. In processes involving a nucleus and an
electromagnetic particle such as a muon, electron, or photon most
activation occurs by exciting the giant dipole resonance through
the exchange of a quasi-real 17 MeV photon. The resonance area is
roughly proportional to the product of the nucleus charge Z and
nucleus atomic number A. The resonance decays predominantly into a
heavy nucleus and a nucleon. If the total resonant cross section
is known, we multiply it by two to obtain an upper limit on total
or partial activities where needed.

According to \cite{nolte1}, the total radionuclide production
cross section is assumed to depend on energy as:
\begin{equation}
\sigma(E)=\sigma_0 E^{0.75}.
\end{equation}
The energy dependence includes secondary production by
bremsstrahlung photons and high energy delta rays.

\begin{figure}
\begin{center}
\epsfig{file=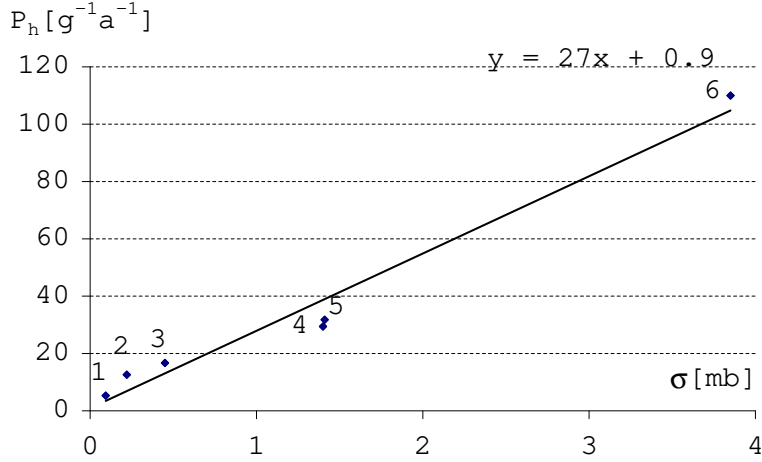,angle=-90, width=10cm}
\end{center}
\caption{ Cross sections for nuclide production by 190 GeV muons
versus rates of nuclide production by cosmic hadrons above
groound. The reactions are labelled as follows:
O$\rightarrow$$^{10}$Be (1), S$\rightarrow$$^{26}$Al (2),
O$\rightarrow$$^{14}$C (3), Ca$\rightarrow$$^{36}$Cl (4),
Si$\rightarrow$$^{26}$Al (5), Fe$\rightarrow$$^{53}$Mn (6).
\label{had-fast}}
\end{figure}

As will be shown in section \ref{resul}, activation by cosmic rays
at sea level is the dominating contribution to our backgrounds.
Since the hadronic spallation cross sections of interest are not
100\% measured \cite{hadron}, we estimated the missing cross
sections using Fig.\ref{had-fast}, which plots the muon-nuclide
cross sections, all known, against the cosmic hadron-nuclide
production rates \cite{hadron}, not all known.

\subsection{Compton absorption length}

The volume-large TPC is surrounded by some 4 mwe of passive
material (see Fig.\ref{drawi}). This material is self-shielded,
meaning that its thickness $L$ is much greater than the Compton
absorption length $\lambda$, and only a very small fraction of the
gamma rays produced within the device enters the active volume.

Under the conditions in which the inner shield is self-shielded,
yet is thin compared to the dimensions of the TPC, the fraction
$f$ of gamma rays entering the TPC, can be estimated analytically:
\begin{equation}
f={1\over 2L}\int_0^1 d\cos{\theta}\int_0^\infty dx
e^{-x/\lambda\cos\theta} ={\lambda\over 4L}
\end{equation}
where $x$ is the distance between the origin point of the emitted
photon and a TPC wall and $\theta$ is the emission direction angle
with respect to the perpendicular of this wall.

The factor of 1/2 in front of the right hand side of the equation takes into
account those gamma rays that point away from the TPC. The result, by no
means new, is displayed for two reasons: first, it makes very clear that
radiopurity near the TPC is of paramount importance, and second, it allows
for a quick cross check of more complex simulations by reducing the
backgrounds to those originating from a smaller, known fraction of the shield.

\begin{figure}[pthb]
\centerline{\psfig{file=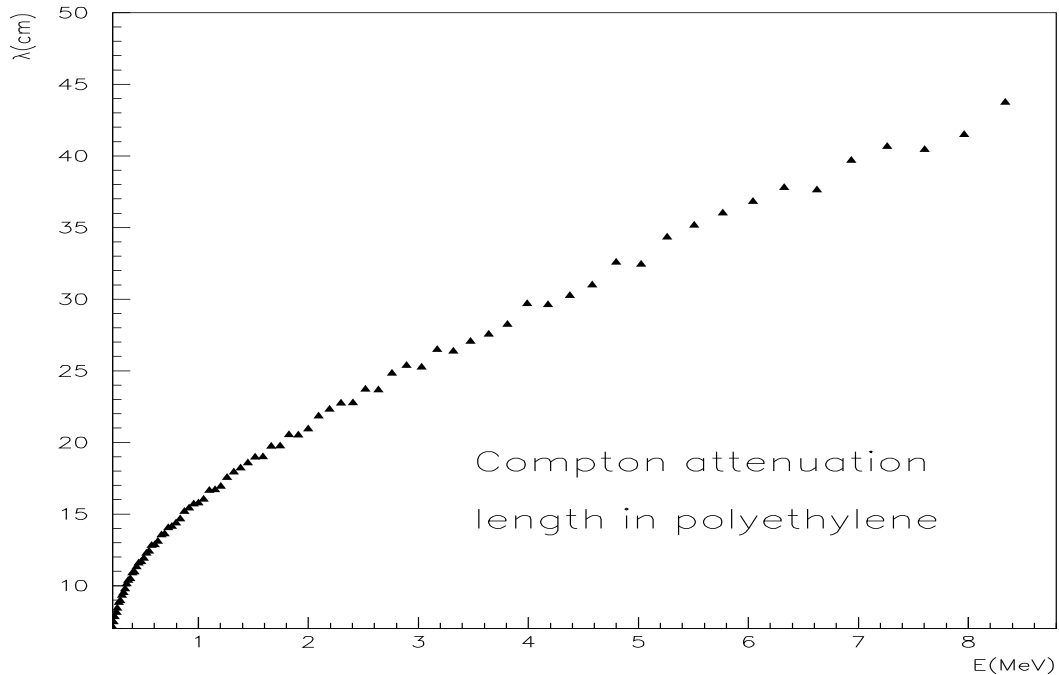,height=3.5in,width=5.5in,angle=0}}
     \caption{GEANT Photon Compton absorption length in
polyethylene ($\rho=1$ g/cm$^3$) as a function of photon energy, and
for a photon cutoff of 200 keV, as described in the text.}
     \label{atten}
\end{figure}

The real life attenuation length of gamma rays originating outside
the TPC involves also geometric factors and energy cuts (once a
photon energy falls below 200 keV, it cannot generate a
candidate). They have been calculated with GEANT and are shown
in Fig.\ref{atten}.

\subsection{Radionuclides of interest}

Considering the activation of inner shielding and gas, all
daughters of C, He, and possibly O (if different type of inner
shielding is chosen) need to be considered. No radionuclide has a
half-life exceeding two months, meaning that activation above
ground decays away during any reasonable construction and assembly
time. The light radionuclides and their radioactivity parameters
are shown in table \ref{c-prod}.

\begin{table}
\begin{center}
\begin{tabular}{|c|c|c|c|c|c|} \hline
&&\multicolumn{2}{|c|}{$\gamma$}&\multicolumn{2}{|c|}{$\beta$}\\\cline{3-6}
nuclide&half-life&avg. E ,&intensity&avg. E,&intensity\\
 &&MeV&&MeV&x charge\\\hline\hline
$^6$He&0.8s&&&1.5&-1\\\hline
$^8$He&0.12s&0.98&0.84&4&-1.84\\\hline
$^8$Li&0.8s&&&6.2&-1\\\hline $^7$Be&53.3d&0.48&0.1&&\\\hline
$^{11}$Be&13.8s&2.1&1&1&-1\\&&5&2.13&&\\&&7&2.7&&\\\hline
$^8$B&0.77s&&&6.7&+1\\\hline$^9$C&0.126&2.3&0.4&7&+1\\\hline
$^{10}$C&19.2s&0.72&1&&\\\hline $^{11}$C&20.39m&&&0.386&+1\\\hline
n&0.1ms&2.2&1&&\\\hline
\end{tabular}
\end{center}
\caption{Radionuclides produced in carbon target and their decay
modes. Only long-lived nuclides dominantly produced are
listed. Where the activity is $\beta^+$, two extra photons
of energy 0.511 keV for each $\beta^+$ are always implied.
\label{c-prod}}
\end{table}

For the steel shielding and copper strips, we also need to
consider isotopes with half-lives exceeding several months (see
table \ref{fecu-prod}). These are activated at the surface.
\begin{table}
\begin{center}
\begin{tabular}{|c|c|c|c|c|c|} \hline
&&\multicolumn{2}{|c|}{$\gamma$}&\multicolumn{2}{|c|}{$\beta$}\\\cline{3-6}
nuclide&half-life&avg. E, &intensity&avg. E,&intensity\\
 &&MeV&&MeV&x charge\\\hline\hline
$^{46}$Sc&84d&1&2&0.11&-1\\\hline
$^{47}$Sc&3.3d&0.16&0.6&2&-1\\\hline
$^{48}$V&16d&1.1&2&0.29&+0.5\\\hline
$^{51}$Cr&28d&0.32&1&&\\\hline $^{52}$Mn&5.6d&0.8&1.9&0.24&+0.3\\
&&1.4&1&&\\\hline$^{53}$Mn&3.7e6y&0.6&1&&\\\hline
$^{54}$Mn&312d&.83&1&&\\\hline $^{56}$Co&77d&0.84&1&0.6&+0.19\\
&&1.2&0.8&&\\ &&1.7&0.16&&\\
&&2.5&0.17&&\\\hline $^{59}$Fe&44d&1.1&1&0.11&-1\\\hline
$^{57}$Co&272d&0.12&1&&\\\hline
$^{58}$Co&71d&0.8&1&0.2&+0.15\\\hline
$^{60}$Co$^{5+}$&5.3y&1.2&2&0.1&-1\\\hline
\end{tabular}
\end{center}
\caption{Radionuclides produced in the steel shielding and the
copper strips and grids. Only long-lived nuclides produced
dominantly are listed. Where the activity is $\beta^+$, two extra photons
of energy 0.511 keV for each $\beta^+$ are always implied. \label{fecu-prod}}
\end{table}

\section{Results.}\label{resul}
\subsection{Above ground activation}
There is no long-term above-ground activation of the inner shield
or gas, as all possible daughters have half-lives well below one
year. The two components that may be activated with long-lived
radionuclides are the steel vessel and the copper strips and
grids. We combine the tables of the preceding section to produce
table \ref{b-upgr}, which evaluates the total rate of photons and
background rates from the corresponding components. The above
ground exposure time was chosen to be infinite.

\begin{table}
\begin{center}
\begin{tabular}{|c|c|c|c|}
\hline target&$R_\gamma(t=0)$&$R_b(t=0)$,&$R_b(t=2y)$\\
&day$^{-1}$&day$^{-1}$&day$^{-1}$\\ \hline plastic &  $3.5
\cdot10^5$&3900 & 0.27
\\\hline
 Cu & $8.2 \cdot10^4$& 8000 & 4000 \\\hline
  Fe & $2\cdot 10^7$& 80 & 15 \\ \hline
\end{tabular}\end{center}
\caption{Expected total rate of gamma emission immediately after
installation due to activation of the TPC components by the cosmic
flux above ground, $R_\gamma(t=0)$, and background rate caused by
this emission at the beginning, $R_b(t=0)$, and two years later,
$R_b(t=2y)$. \label{b-upgr}}
\end{table}

We find already the largest background from $^{60}$Co activated in
the copper immediately after being taken underground. A two year
wait cuts the background down to 4000 events per day. However,
copper will be electrodeposited on plastic to produce both strips
and the various HV grids. Electrodeposition will probably strongly
reduce most radio-impurities, including certainly cobalt. It
should also be mentioned that tracks originating in the detector
plane will look anomalous (the track beginning will be recorded by
a single wire or strip, for example).

\subsection{Underground activation: TPC gas}

Table \ref{b-gas} shows the activity in the TPC gas at four depths
varying from 1500 to 3000 mwe. As can be seen, a $^{11}$C
component dominates, however the backgrounds are still a few
orders of magnitude below the expected TPC backgrounds of a few
hundred events per day. Hydrogen is inert from a cosmogenesis
point of view, while He contributes a modest tritium background on
which we do not trigger.

\begin{table}
\begin{center}
\begin{tabular}{|c|c|c|}
 \hline
  depth, & isotope & $R_{\beta}$,\\
  mwe &  & day$^{-1}$\\\hline\hline
    &$^6$He&    0.029  \\\cline{2-3}
    &$^8$He&    0.0038  \\\cline{2-3}
    &$^{11}$Be& 0.0041  \\\cline{2-3}
1500    &$^8$B& 0.013  \\\cline{2-3}
    &$^9$C& 0.0085  \\\cline{2-3}
    &$^{11}$C&  1.6  \\\cline{2-3}
    &total& 1.7
\\\hline
2000    &total& 0.53\\\hline 2500    &total& 0.2
\\\hline
3000    &total& 0.083\\\hline

\end{tabular}
\end{center}
\caption{Background rate due to activation of methane in the gas
by fast muons. Since the photon interaction probability is small,
only $\beta$-emitters are included.\label{b-gas}}
\end{table}

\subsection{Underground activation: inner shield}
Table \ref{b-plast} shows the activity in the inner shield at four
depths varying from 1500 to 3000 mwe. The table considers the
cosmic rays which enter the TPC (all the short-lived radionuclides
are removed).

The cosmic rays whose total induced
activity needs to be considered are those which
do not enter the TPC and are not vetoed. Their total contribution to the
background is related via Eq. (8) to the number of cosmic rays which
cross the inner $\lambda$/4-thick layer of plastic but avoid the TPC.
These are approximately 0.2\% of the total cosmic rays
crossing the TPC. We conclude that
the contribution from non-vetoed short-lived species is
negligibly small.
\begin{table}
\begin{center}
\begin{tabular}{|c|c|c|c|c|}
 \hline
  depth, & isotope & $R_{emit}$,& $R_{ent}$,&$R_{int}$,\\
  mwe &  & day$^{-1}$&day$^{-1}$ &day$^{-1}$\\\hline\hline
&$^8$He&$<$ 7.4 &$<$    0.19  &$<$    0.016   \\\cline{2-5}
&$^7$Be& 80   &   1.6    &   0.17  \\\cline{2-5} &$^{11}$Be&$<$ 40
&$<$ 2.2    &$<$    0.067   \\\cline{2-5} &$^8$B& 49 & 0.97 & 0.11
\\\cline{2-5} 1500&$^9$C& 58 & 0.91 &   0.085
\\\cline{2-5} &$^{10}$C& 380   &   9    & 0.83\\\cline{2-5}
&$^{11}$C& 6300 &   130 &   13
\\\cline{2-5} &total& 7000 & 140  & 15 \\\cline{2-5} &n& 3900  &
140  & 3.3
 \\\hline
2000&total& 2200  &   45   &   4.8   \\\hline 2500&total& 850 & 17
&   1.8   \\\hline 3000&total& 350   & 7.1 & 0.75
\\\hline
\end{tabular}
\end{center}
\caption{Total rate of emitted photons due to activation of the
plastic inner shield by fast muons. The labelling is following:
$R_{emit}$ is the rate of emitted photons, $R_{ent}$ the number of
photons entering the TPC and $R_{int}$ the number of photons
interacting with the TPC gas. All short-lived radionuclides are
removed. \label{b-plast}}
\end{table}
\subsection{Underground activation: copper and steel}
Table \ref{b-fecu} shows the activity in the copper and the steel.
As can be seen, this contribution is much smaller than from the
inner shield.
\begin{table}
\begin{center}
\begin{tabular}{|c|c|c|c|c|}
\hline
  depth, mwe & 1500 & 2000& 2500& 3000 \\\hline
  $R_{Fe}$, day$^{-1}$ &  $1.9\cdot10^{-3}$&$6.6\cdot10^{-4}$  & $2.3\cdot10^{-4}$ & $3.5\cdot10^{-5}$ \\ \hline
  $R_{Cu}$, day$^{-1}$ & 0.063 & 0.021 & 0.0078 & 0.0032 \\ \hline
\end{tabular}
\end{center}
\caption{Cosmogenic background caused by fast muons in the steel
shielding and the copper strips and grids.\label{b-fecu}}
\end{table}

\subsection{Underground activation: muon bremsstrahlung and muon capture}

The energy deposited by fast muons in our device increases by a
factor of 1.7, to $10^5$ GeV per day (1500 mwe), when
bremsstrahlung is taken into account. At 3000 mwe, the factor and
deposited energy are 2.2 and $10^4$ GeV per day, respectively. The
average photon energy (energy weighted) is approximately 100 (at
1500 mwe) to 150 GeV (at 3000 mwe). The average photon energy
above 200 keV varies from 15 (at 1500 mwe) to 20 GeV (at 3000
mwe). Because of the $Z^2$ dependence of bremsstrahlung, about
97\% of the photons are emitted in the iron.

The total number of photons between 200 keV and 10 MeV is about
1000 (at 1500 mwe) to 50 (at 3000 mwe) per day, and they are emitted
nearly parallel to the muon. In practice, simulations showed that
this source of backgrounds contributes less than 0.1 candidate
events per day at any depth.

Muon capture at these depths ranges from 0.03/ton/day (1500 mwe)
to 0.002/ton/day (3000 mwe). These numbers are at least two orders of
magnitude below those provided in table \ref{reac-prob} and are
negligible.

\section{Cross checks and conclusions}\label{x-check}
\subsection{Cross checks}
There are ways to roughly cross check our
results. If radiopurity is expressed in $^{238}U$ equivalent
purity, BOREXINO purity needs are about $10^{-16}$  g/g
\cite{borexino} and less than ten events per day. In the case of
the TPC, the cage needs to have a purity of $10^{-13}$ g/g
\cite{position} and less than 600 events per day (the two numbers
do not scale the same way because  the cage weighs less than the
BOREXINO inner detector, and because only a fraction of the gammas
originating in the cage will interact in the TPC volume).

If BOREXINO would observe negligible backgrounds at 4500 mwe (1200 mwe below
its current location), then the TPC will observe negligible backgrounds
at roughly 1200 mwe, if scaled by purity, and at 2000 mwe, if scaled by
number of events.

\subsection{Conclusions}\label{conclu}
The TPC has a unique set of properties amongst solar neutrino detector.
\begin{enumerate}
\item the active volume occupies 62\% of the total device, but weighs
only 0.1\% of the total weight. The active volume intercepts a
large fraction of the cosmic rays and most short-lived
radionuclides are simply vetoed by the observable muon track.
\item it distinguishes events of multiplicity two or more from events of
multiplicity one.
\item its active volume is virtually free of any significant
cosmogenesis backgrounds. Cosmogenesis backgrounds come from
the surrounding materials, which are self-shielded.
\item the TPC directional capability further eliminates most events not
pointing at the sun, and subtracts statistically the remainder.
\end{enumerate}
All of the characteristics above combine to give a detector
virtually immune of cosmogenesis backgrounds at depths as low as
1500 mwe.

The largest long-term backgrounds found by this analysis were from
copper and steel activated at the surface (which are independent
of depth). Two lessons are to be learned from this analysis.
First, copper will have to be electrodeposited, a procedure known
to suppress backgrounds by at least a few orders of magnitude. In
particular, electrodeposition virtually eliminates $^{60}$Co
\cite{brodzi}. Second, during the R\&D phase, there will be of
order 0.5 meters of inner shielding, and the backgrounds should be
dominated by $^{54}$Mn (after the R\&D device is adequately
shielded from the surrounding rock). There will be a premium in
obtaining steel that has been "sheltered" from cosmic radiation.
In the final detector, $^{54}$Mn is taken care of by increasing
the thickness of the inner shield by 0.5 meters (originally
\cite{position} a one meter thick inner shield was considered).

In practice, the TPC depth is not limited by backgrounds, but
rather by dead time, surface charging and future applications for
the TPC. Civil engineering needs to be assessed too to evaluate
the cost of a very large hall as a function of depth.

As discussed above, each cosmic event vetoes 1.8\% of a second, therefore it
is advisable
not to build the TPC above the depth where the cosmic muon rate is 1 Hz
(approximately 1900 mwe).

Surface charging has been discussed in  \cite{position}. The
advantages in building a TPC with insulating walls are better
electrostatic stability and better control of the materials. The
exact limit depends on the final choice of plastic and the safety
factor above the charge-up limit. For acrylic it is probably
advisable not to go above 2000 mwe, whereas a plastic as conductive
as polyethylene would not be charged up under any realistic
circumstances.

Future applications of the TPC may include dark matter searches,
where low-energy nuclear recoil (roughly 20 keV) is the signature
for dark matter interactions. The signature is non-directional,
and the recoil energy is proportional to the nuclear mass of the
target. In practice one tries to see a change in the very low
energy spectrum when the gas nuclear mass changes ($e.g.$, varying
the gas from helium to carbon, then to neon, argon, xenon). The
lack of directional information, plus varying cosmogenesis
backgrounds, will perhaps provide the most stringent constraints
on depth.

\subsection{Acknowledgements}
We are particularly indebted to R. Brodzinski and E. Nolte. Dr. Brodzinski
pointed out that backgrounds would probably be dominated by above ground
activation, and provided information and criticism that was helpful
throughout the paper. Prof. Nolte provided an early version of his
soon-to-be published paper which provided the backbone for all the
calculations and results presented in this paper.

\end{document}